\documentclass[onecolumn, aip]{revtex4-1}

\usepackage{amsmath}
\usepackage{graphicx}
\usepackage{epstopdf}

\begin{document}

\title{Damage threshold and focusability of mid-infrared free-electron laser pulses gated by a plasma mirror 
with nanosecond switching pulses}

\author{Xiaolong Wang}
\noaffiliation
\author{Takashi Nakajima}
\email[]{electronic mail: nakajima@iae.kyoto-u.ac.jp}
\noaffiliation
\author{Heishun Zen}
\noaffiliation
\author{Toshiteru Kii}
\noaffiliation
\author{Hideaki Ohgaki}
\email[]{electronic mail: ohgaki@iae.kyoto-u.ac.jp}
\noaffiliation
\affiliation{Institute of Advanced Energy, Kyoto University, Gokasho, Uji, Kyoto 611-0011, Japan}

\date{\today}

\begin{abstract}
The presence of a pulse train structure of an oscillator-type free-electron laser (FEL)
results in the immediate damage of a solid target upon focusing. 
We demonstrate that the laser-induced damage threshold can be significantly improved 
by gating the mid-infrared (MIR) FEL pulses with a plasma mirror. Although the switching 
pulses we employ have a nanosecond duration which does not guarantee the clean 
wavefront of the gated FEL pulses, the high focusablity is experimentally confirmed 
through the observation of spectral broadening by a factor of 2.1 when we tightly 
focus the gated FEL pulses onto the Ge plate.
\end{abstract}


\maketitle

FELs \cite{Deacon:1977zz} are powerful coherent light sources in a wide spectral range 
covering from the extreme ultraviolet to the microwave wavelength regions 
\cite{OShea:2001gv}. 
The FELs in the MIR region are of particular interest for molecular spectroscopy, 
because the frequency of molecular vibrations often fall in this wavelength region. 
Most of the FELs in MIR, which include Kyoto University free-electron laser (KU-FEL)  
\cite{Ohgaki:2008}, are an oscillator-type and they usually have a dual-pulse structure: 
A macropulse duration is about millisecond to microsecond in which thousands of 
micropulses with sub-picosecond durations are sitting with a few hundred picosecond 
to several nanosecond time interval between successive micropulses. 

For some applications of FELs which require the time-resolution, high peak intensity, 
etc., the time-gating and high focusability of FEL pulses without damaging a target 
would be very important. 
In this paper we demonstrate that these two requirements for the FEL pulses can be 
fulfilled with the aid of a plasma mirror \cite{Alcock:1979wt, Knippels:1998, Thaury:2007ua}. 
There are two new features in the present work. The first one is a use of switching pulses 
with a {\it nanosecond} duration for a plasma mirror, and the second one is a 
demonstration of high focusability through the observation of spectral broadening 
for a {\it pulse train} with the aid of a plasma mirror. 

The former one is somehow surprising if we recall that the plasma mirror is 
conventionally switched by short laser pulses to avoid the wavefront distortion 
of gated pulses so that the high focusability is guaranteed. 
The reason for this is clear if we recall that the surface plasma induced by the 
switching pulse must exceed the critical density well before its thickness 
expands more than the dimension of the incident laser wavelength. 
Otherwise a wavefront distortion of the gated incident pulse might take place. 
Mathematically it can be expressed by the inequality, $c_{s} \Delta t < \lambda$, 
as proposed in Ref. 7 where $c_s$ is a sound speed, $\Delta t$ is 
time from plasma formation to the peak of the pulse, and $\lambda$ is a wavelength 
of laser pulses to be gated. This inequality physically makes sense, but our results 
seem to imply that it is perhaps a sufficient condition to have a clean wavefront, 
since the violation of this inequality does not immediately mean that we necessarily 
suffer from the wavefront distortion and lose focusability. 
This implies that the use of nanosecond switching pulses for a plasma mirror 
would have a practical significance for some cases because of the simple and
cost-effective experimental configuration, in particular when an expensive ps or fs
laser is not available to switch the plasma mirror. 

As for the latter, the high focusability of the gated FEL pulses allows us to reach
the high peak intensity at which nonlinear optical effects take place. 
Spectral broadening induced by self-phase modulation (SPM) in nonlinear media 
\cite{Agrawal:1989ue, Corkum:1985uj, Ashihara:2009uy} is one of such effects. 
To our knowledge, however, all the results of spectral broadening are obtained 
for {\it isolated} laser pulses. If we focus several thousands of MIR pulses 
with a short time interval onto the nonlinear medium (Ge plate) to induce SPM, 
the Ge plate is very easily damaged by a single shot. 
As one can easily imagine the accumulated damage on the Ge plate by a {\it burst} 
of laser pulses with a short time interval is much more severe than those by the 
same number of isolated laser pulses. 
For the case of KU-FEL a burst of $>$ 4000 micropulses with 0.6 ps duration 
\cite{Qin:2013}, $\sim\mu$J energy, and 350 ps time interval interacts with  
the Ge plate if we do not employ the plasma mirror. 
With a plasma mirror we can temporally select only a small number 
of micropulses (only $\sim$15 out of $>$ 4000 micropulses in a single macropulse) 
from KU-FEL and induce SPM without damaging the Ge plate even when 
the pulse is very tightly focused. 
Thus the use of a plasma mirror enables us to improve the damage threshold 
(in terms of the peak intensity) of the Ge target by at least 10 times and realize 
high intensity without damaging the target. 
Note that the damage threshold is usually defined in terms of fluence. 
Such a definition, however, loses its meaning if we wish to see the nonlinear
phenomena. Defining the damage threshold in terms of peak intensity would be 
more appropriate. 
After the optimization the spectral bandwidth is increased by a factor of 2.1,
which provides us with a more spectral range for spectroscopic applications of 
KU-FEL. 

\begin{figure}
\centerline{
\includegraphics[width=8cm]{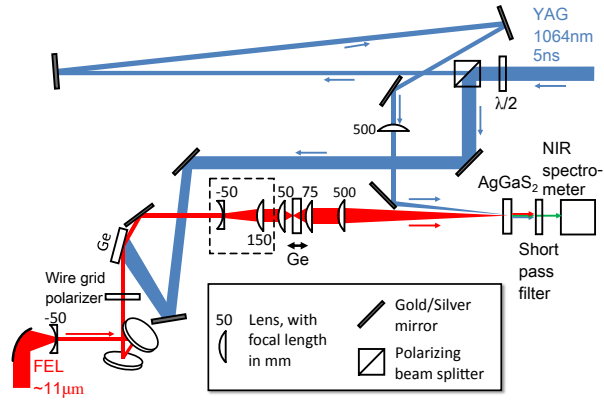}
}
\caption{
(Color online) 
Experimental setup. The first Ge plate serves as a plasma mirror
when irradiated by 1064 nm pulses, while the second Ge plate serves as a nonlinear 
material to induce SPM. When the 1:3 beam expander in the dashed box is removed,
the diameter of the beam waist is increased by a factor of $\sim$3.
}
\label{fig:experiment_plan}
\end{figure}

The experimental setup is shown in Fig. \ref{fig:experiment_plan}. The 11 $\mu$m FEL 
beam with a macropulse energy of 10 mJ operated at 1 Hz repetition rate 
is shrunk to the beam size of 1/6 from the original one by an off-axis parabolic mirror 
of 152.4 mm effective focal length and a plano-concave lens of -25.4 mm focal length. 
A periscope and a BaF$_{2}$ wire-grid polarizer (Thorlabs WP25H-B), respectively, 
changes and purifies the polarization of the FEL beam to the horizontal direction 
(p-polarization with respect to the first Ge plate (Thorlabs WG91050, uncoated) 
which serves as a plasma mirror) to minimize the surface reflection at the Brewster's 
incident angle which is $76^{\circ}$ for the Ge plate at 11 $\mu$m wavelength. 

\begin{figure}
\centerline{
\includegraphics[width=7cm]{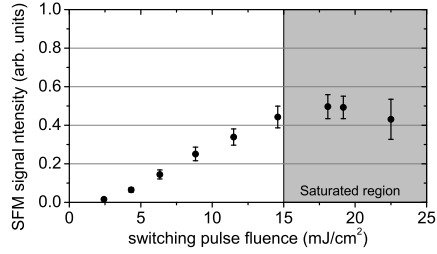}
}
\caption{
SFM signal intensity as a function of 1064 nm switching pulse fluence. 
The vertical scale is normalized so that the SFM signal becomes unity when the 
first Ge plate is replaced by a normal Au mirror, and hence it may be read 
as a reflectivity. 
}
\label{fig:SFM_fluence}
\end{figure}

A Nd:YAG laser pulse (Rayture systems GAIA-III, 1064 nm, 5 ns pulse duration, 
80 mJ pulse energy) is synchronized with a RF source of KU-FEL via a delay generator 
(SRS DG645) so that the irradiation timing of the 1064 nm pulse onto the Ge plate 
and the arrival timing of the peak of the FEL macropulse at the Ge plate coincide. 
The 1064 nm pulse is split into two by a $\lambda$/2 plate and a polarizing 
beam splitter cube (CVI Melles Griot PBSH-450-1300-050). We can control the 
branching ratio of the pulse energy in two pulses by adjusting the angle of the 
$\lambda$/2 plate. 
The most energy of the 1064 nm pulse is used to activate the plasma mirror, and 
the rest is used for sum-frequency mixing (SFM) as we will explain later. 
This way we are able to temporally select a small number ($\sim15$) of FEL 
micropulses out of $>$ 4000 micropulses within the same macropulse as a reflected 
portion by the plasma mirror. The gate width of the plasma mirror is approximately 
the same with the duration of the 1064 nm pulse, which is 5 ns. 
In Fig. \ref{fig:SFM_fluence} we plot the change of the SFM signal intensity as a 
function of 1064 nm pulse fluence onto the first Ge plate. Since the vertical scale 
is normalized so that the SFM signal becomes unity when the Ge plate is replaced 
by a normal Au mirror, it may be read as the reflectivity of the plasma mirror. 
As the 1064 nm pulse fluence increases the density of the surface plasma increases 
and the reflectivity also increases up to about 0.5. After it reaches the maximum 
it stays almost at the same value. 
Note that the SFM signals in Fig. \ref{fig:SFM_fluence} are normalized by 
the energies of the incident FEL macropulse before the plasma mirror and 
the 1064 nm pulse used for SFM, and hence the error bars represent the
repeatability of the plasma mirror operation.

\begin{figure}[b]
\centerline{
\includegraphics[width=8cm]{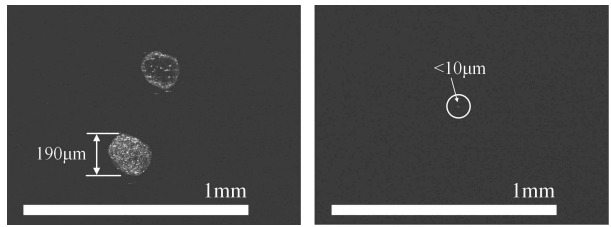}
}
\caption{
Comparison of the surface images of the second Ge plate after the irradiation 
of (a) a single FEL macropulse (which contains $\sim$4000 micropulses) at 
$<$ 6 GW/cm$^{2}$ without the plasma mirror and (b) more than 3600 FEL 
macropulses (each of which contains $\sim$15 micropulses) at 
$\sim$60 GW/cm$^{2}$ with the plasma mirror. 
}
\label{fig:Ge surface}
\end{figure}

We emphasize that the use of a plasma mirror is the key to avoid the damage of the 
second Ge plate in which the SPM takes place. 
A comparison of the surfaces of the second Ge plate (Thorlabs WG91050-F, 
AR-coated, thickness 5 mm) with/without the plasma mirror is presented in 
Fig. \ref{fig:Ge surface}. Although the origin of small spots in Fig. \ref{fig:Ge surface}(b) 
is unidentified, we believe they were not produced by the tightly focused FEL pulses, 
since the SFM signal intensity remained the same during the irradiation of tightly 
focused FEL pulses for more than an hour. 
The lifetime of the plasma mirror for our specific case is at least several hours
even without translating the position of the second Ge plate. 
It is already so long under such conditions, because the required plasma density 
for the plasma mirror is rather low for MIR incident light and therefore the Ge 
surface is hardly damaged by the 1064 nm switching pulses: 
Recall that the critical plasma density to activate the plasma mirror is 
proportional to the square of the incident FEL frequency \cite{Alcock:1979wt}. 
For instance the critical plasma density to activate the plasma mirror 
for the 11 $\mu$m pulse is approximately two orders of magnitude smaller 
than that for the 1 $\mu$m pulse.
Needless to say, if we introduce a translation mechanism of the Ge plate
to use a fresh Ge surface for every hour, the lifetime can be far longer than 
several hours.

% the loose and tight focusing conditions: 
%Without/with a 3x telescope consisting of a $f=$50 and 150 mm lenses 
%(dashed box in Fig. \ref{fig:experiment_plan}) before the second Ge plate, which we call 
%loose and tight focusing conditions, the ratio of the spot sizes 
%(peak intensities) is 9:1 (1:9). If we do not use the plasma mirror the second Ge plate 
%is easily damaged by a single shot under the loose focusing condition. 
%In contrast, when we employ the plasma mirror the second Ge plate is free from damage 
%even under the tight focusing condition after the irradiation of several hours. 

The FEL micropulses gated by the plasma mirror are now focused onto the second 
Ge plate with a 1:3 beam expander before the $f=$50 mm focusing ZnSe lens. 
Assuming that the FEL pulses have Gaussian transverse profiles, the diameter 
at the focus is 45 $\mu$m. 
Because the reflectivity of the plasma mirror is $\sim$50 \% and the 
transmission of each ZnSe lens is 98 \%, we estimate the peak intensity of the 
gated micropulses at the focus is $\sim$60 GW/cm$^{2}$, which has to be 
cross-checked since the FEL pulses after the plasma mirror may not have 
a good focusability.  
Note that the tight focusing leads to the Rayleigh range as short as 0.15 mm. 

We detect the FEL spectra after passing through the second Ge plate by the 
SFM technique \cite{Wang:2012wa} : 
The FEL pulses at 11 $\mu$m are upconverted into the near-infrared (NIR) pulses 
at 970 nm by mixing with 1064 nm pulses in a AgGaS$_{2}$ crystal 
(type I, cut angle $37^{\circ}$, thickness 0.5 mm), 
and we measure the NIR pulse spectra with a compact CCD spectrometer. 
The setup is very similar to that described in our recent paper \cite{Wang:2012wa} 
with a few differences that the spectrometer (Ocean Optics Maya2000 Pro-NIR) 
we employ now has much better sensitivity and spectral resolution than the 
previous one.  
Once the NIR pulse spectra are obtained we can easily retrieve the FEL spectra
as we have described in Ref. \cite{Wang:2012wa}. 
The AgGaS$_{2}$ crystal thickness we employ here is 0.5 mm instead of 2 mm 
for the previous work to have 4$\times$ broader phase-matching bandwidth, 
$\sim$3 $\mu$m. 

\begin{figure}
\centerline{
\includegraphics[width=7cm]{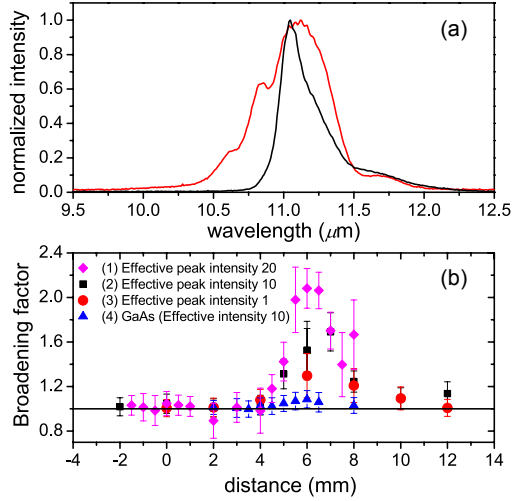}
}
\caption{
(Color online) 
(a) The FEL spectra before (black) and after (red) the spectral broadening 
at the optimal position of the second Ge plate. The linewidth (FWHM) of the
spectrum is increased by a factor of 2.1.
(b) Bandwidth of the FEL spectra as a function of position of the nonlinear
material (Ge or GaAs), where (1) the optimal broadening obtained with maximum
FEL pulse energy, (2) and (3) comparison of tight and loose focusing conditions
when the FEL pulse energy is fixed at half the maximum, and (4) broadening 
obtained with a GaAs crystal in place of Ge, with half maximum pulse energy
and tight focusing.
Note that ``zero'' of the horizontal axis does not represent 
the exact focus position but a relative quantity due to the difficulty to 
determine ``zero''. 
}
\label{fig:broadening_distance}
\end{figure}

For a given nonlinear material and wavelength of the incident pulse, the strength of 
SPM process depends on the product of peak intensity of the incident pulse and 
effective propagation length in the nonlinear material. 
Although other nonlinear effects such as self-focusing and nonlinear absorption may 
alter the effective propagation length from the rough estimation, this length should not 
be very different from the Rayleigh range, since the peak intensity of the FEL 
micropulse is not very high compared to femtosecond pulses conventionally employed
in SPM process. 
By recalling that the peak intensity of the micropulse is inversely proportional to the  
square of the beam diameter, and the Rayleigh range is proportional to the beam 
diameter at the waist, 
the use of a pulse with a smaller diameter, or tighter focusing, should induce more 
spectral broadening. To confirm this, we have measured the FEL spectra under the 
tight (with the beam expander) and loose (without the expander) focusing conditions, 
with all the other parameters kept to be the same. 
The results are shown in Fig. \ref{fig:broadening_distance}. 
It clearly shows that the tight focusing is more favorable than the loose focusing, and 
the broadening factor is 1.7 (squares in Fig. \ref{fig:broadening_distance}(b)) by the 
former compared to 1.3 by the latter (circles in Fig. \ref{fig:broadening_distance}(b)).
When we further increase the beam energy under the tight focusing condition, 
the broadening factor becomes 2.1 (diamonds in Fig. \ref{fig:broadening_distance}(b)). 
The increment of the broadening factor from 1.7 to 2.1 as a result of doubling 
the incident pulse energy turns out to be rather marginal. It is likely that the nonlinear 
absorption plays some role to reduce the pulse energy (and hence effective peak 
intensity) before the pulse reaches the peak value. If the micropulse duration
were shorter the nonlinear absorption would play a lesser role under the similar
condition. 

We have also taken the similar data using a GaAs crystal to find that the spectral
broadening is almost negligible (triangles in Fig. \ref{fig:broadening_distance}(b)). 
Note that the bandgap of GaAs (1.43 eV) is about twice as large as that of 
Ge (0.67 eV). 
Although a GaAs crystal was found to be a good choice to broaden the femtosecond 
5 $\mu$m pulses \cite{Ashihara:2009uy}, this is not the case for the 11 $\mu$m 
pulses and Ge has been found to be more effective.

\begin{figure}
\centerline{
\includegraphics[width=8cm]{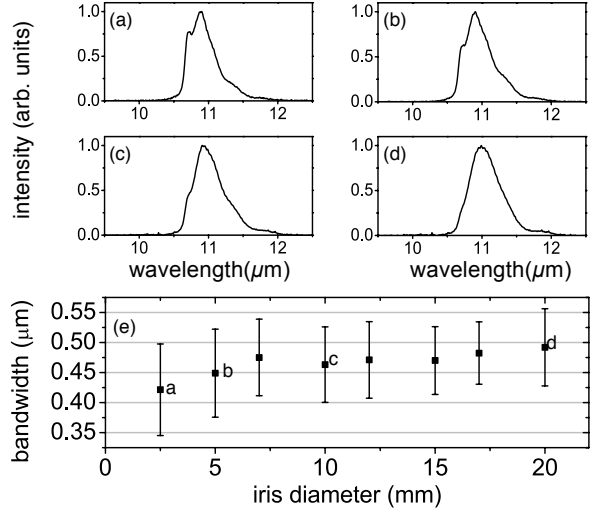}
}
\caption{
FEL spectra after the second Ge plate transmitted through an iris
with different diameters, (a) 2.5, (b) 5, (c) 10, and (d) 20 mm. 
(e) Change of the bandwidth of FEL pulses transmitted through an iris with 
different diameters. 
}
\label{fig:aperture}
\end{figure}

If we may assume that the FEL micropulses have Gaussian temporal profile given by 
$I(t)=I_{p}\exp(-2 t^{2}/\tau^{2})$, the maximum frequency shift can be estimated to be 
$|\Delta\omega(t)|_{\text{max}}=
|\Delta\omega(\tau/2)|\sim0.5|\Delta\phi|_{\text{max}}\Delta\omega_{0}$, 
in which $|\Delta\phi|_{\text{max}}=(\omega/c)n_{2}I_{p}L$ is the maximum phase shift, 
$I_{p}$ and $L$ are the peak intensity and effective propagation length in the nonlinear
material, and $\Delta\omega_{0}$ is the bandwidth of the incident FEL pulses 
\cite{Ashihara:2009uy}. 
By substituting the nonlinear refractive index $n_{2}=9900 \times 10^{-16}$ cm$^{2}$/W 
for Ge and the experimentally measured broadening factor, 2.1, we can estimate the 
product of $I_{p}$ and $L$ to be $I_{p}L$=600 MW/cm. 
If we further assume that $L$ may be approximated by the Rayleigh range which is 
0.15 mm under the tight focusing condition, we find that the peak intensity is 
$I_{p}$=43 GW/cm$^{2}$. 
This value agrees well with the previously estimated peak intensity, 60 GW/cm$^{2}$, 
from the formula of focusing for a Gaussian beam. 

We also measure the FEL spectra transmitted through the different diameters
of iris after the second Ge plate (see Fig. \ref{fig:experiment_plan}) under the tight 
focusing condition. The results are shown in Fig. \ref{fig:aperture}. 
Although the measured spectra with various iris diameters are not very different 
we can still see some small differences in Figs. \ref{fig:aperture}(a)-\ref{fig:aperture}(d): 
Regarding the spectral shape the slopes are a bit sharper for the inner part of the beam
and the shoulder structure is more pronounced (Fig. \ref{fig:aperture}(a)).  
This feature gradually disappears as the iris diameter becomes larger 
(Figs. \ref{fig:aperture}(a)-\ref{fig:aperture}(d)). 
A similar tendency has been observed by Ashihara {\it et al.} for the femtosecond 
5 $\mu$m pulses broadened with a GaAs crystal \cite{Ashihara:2009uy}, and 
it is interpreted as an interplay of SPM and parametric processes during propagation
in a nonlinear material. 
As for the spectral width with different iris diameters, it turns out that spectra 
for the inner part is slightly narrower (Fig. \ref{fig:aperture}(e)). 
This may be because of the formation of the plasma which works to self-defocus 
the beam at the center, and accordingly lowers the intensity at the center. 
Further investigation is necessary to clarify this point. 

\vspace{10pt}

In summary, we have demonstrated that the plasma mirror switched by nanosecond 
pulses works well not only to temporally select MIR FEL micropulses but also to realize 
tight focusing up to tens of GW/cm$^{2}$ without damaging the target. 
Because $\sim$GW/cm$^{2}$ is the intensity range above which the various 
nonlinear phenomena set in for solid targets, to be able to go beyond this intensity 
range without damaging the solid target using a simple and cost-effective experimental 
configuration would have a practical significance. 
The high focusability of the gated FEL pulses has been confirmed through the observation 
of spectral broadening by a factor of 2.1, which implies that the peak intensity of 
40-60 GW/cm$^{2}$ has been achieved. 
In the near future we plan to apply the developed technique to carry out the
time-resolved study of a solid target. 

\vspace{10pt}

TN acknowledges fruitful discussions with Prof. Satoshi Ashihara. 
This work was supported by a grant-in-aid for scientific research from the Ministry of 
Education and Science of Japan.

\end{document}